\documentclass[a4paper,11pt]{article}
\pdfoutput=1 

\usepackage{jheppub}
\usepackage[normalem]{ulem}

\usepackage{dcolumn}
\usepackage{slashed}
\usepackage{bm,amssymb,slashed,graphicx,multirow,soul,mathtools,xspace,array,tikz,amsmath,siunitx}
\usepackage[compat=1.1.0]{tikz-feynman}   
\usepackage{float}
\usepackage[colorlinks=true,citecolor=blue,urlcolor=blue,linktocpage=true,
linkcolor=blue]{hyperref} 
\allowdisplaybreaks
\usepackage{ bbold }
\usepackage{braket}
\usepackage{multirow}

\definecolor{nicered}{rgb}{0.7,0.1,0.1}
\definecolor{nicegreen}{rgb}{0.1,0.5,0.1}
\definecolor{violet}{rgb}{0.7,0.3,0.3}
\newcommand{\lp}{\left(}
\newcommand{\rp}{\right)}

\newcommand{\beq}{\begin{equation} }
\newcommand{\eeq}{\end{equation}} 
\newcommand{\bi}{\begin{itemize} }
\newcommand{\ei}{\end{itemize} }

\definecolor{Red}{rgb}{1.,0.,0.}
\definecolor{Grn}{rgb}{0.,0.75,0.}
\definecolor{Blu}{rgb}{0.,0.,1.}
\definecolor{Pink}{rgb}{1,0.08,0.58}

\usepackage[T1]{fontenc} 

\usepackage{amsmath,amssymb,color,slashed}
\allowdisplaybreaks  

\title{Direct CKM determination from W decays at future lepton colliders}

\author[1]{David Marzocca}
\author[2]{, Manuel Szewc}
\author[3]{, Michele Tammaro}

\affiliation[1]{INFN, Sezione di Trieste, SISSA, Via Bonomea 265, 34136, Trieste, Italy}
\affiliation[2]{Department of Physics, University of Cincinnati, Cincinnati, Ohio 45221, USA}
\affiliation[3]{INFN Sezione di Firenze, Via G. Sansone 1, I-50019 Sesto Fiorentino, Italy}

\emailAdd{david.marzocca@ts.infn.it}
\emailAdd{szewcml@ucmail.uc.edu}
\emailAdd{michele.tammaro@fi.infn.it}

\begin{document}

\abstract{
    We project the reach of future lepton colliders for measuring CKM elements from direct observations of $W$ decays. We focus our attention to $|V_{cs}|$ and $|V_{cb}|$ determinations, using FCC-ee as case study. We employ state-of-the-art jet flavor taggers to obtain the projected sensitivity, and scan over tagger performances to show their effect. We conclude that future lepton collider can sizeably improve the sensitivity on $|V_{cs}|$ and $|V_{cb}|$, albeit the  achievable reach will strongly depend on the level of systematic uncertainties on tagger parameters.
}

\maketitle

\flushbottom


\section{Introduction}
\label{sec:Introduction}
The Cabibbo-Kobayashi-Maskawa (CKM) matrix describes transitions between up and down quark mass eigenstates in weak interactions.
The theoretical and experimental establishment of quark mixings marked one of the milestones of the Standard Model (SM) of particle physics. Five decades of efforts lead to pinning down the size and phases of CKM elements with a precision ranging from sub per-mille to few percent~\cite{ParticleDataGroup:2022pth}.
The knowledge of the absolute size of CKM matrix elements allows to test the unitarity of the CKM matrix, a crucial prediction of the SM. These values also appear as inputs in rare flavour-changing processes and can be the limiting factor in the sensitivity to possible physics beyond the SM.
Currently, CKM matrix elements are extracted from measurements of leptonic and semileptonic decays of hadrons, hadronic $\tau$ decays, or studies of meson mixing. In all these cases a precise knowledge of non-perturbative hadronic matrix elements is required, which is often the limiting factor for the achievable precision.

A complementary measurement of CKM matrix elements, free from hadronic matrix elements uncertainties, can be performed at future lepton colliders, such as FCC-ee, ILC, and CepC, from the direct on-shell production of $W$ bosons and subsequent decay in two quark jets \cite{Charles:2020dfl,deBlas:2024bmz}. The decay width of the process $W^+\to u_i \bar d_j$, as well as its conjugate, is directly proportional to the CKM element $|V_{ij}|^2$. Colliding electrons and positrons at center of mass energies close to the $WW$ production threshold, these colliders will provide a large sample of events in an exquisitely clean environment. Additionally, state-of-the-art jet flavor taggers, calibrated precisely thanks to the $\approx 10^{12}$ $Z$ bosons produced in the $Z$-pole run,  allow to reconstruct the initial quark flavors of an event with high efficiency and relatively low mistag probabilities.
For the remainder of this study we focus on the projected performances of jet-flavor taggers at the FCC-ee future collider, obtained with the IDEA detector concept~\cite{Bedeschi:2022rnj, Gouskos:taggers, Selvaggi:taggers}, and perform our analysis based on these values. However, we expect similar results to apply also for CepC~\cite{CEPCPhysicsStudyGroup:2022uwl, Liang:2023yyi}. At the ILC, while a similar number of $W$ bosons can be expected, the lack of a Tera-$Z$ run does not allow to reach the same level of precision in the jet tagging performance \cite{ILDConceptGroup:2020sfq}. 
A measurement of this kind for $|V_{cs}|$ was performed at LEP2~\cite{DELPHI:1998hlc}, reaching a precision of $\sim 30\%$.
The measurement of the total hadronic branching ratio of the $W$ boson, together with precise SM predictions, can also be used to extract a value for $|V_{cs}|$ with an uncertainty of approximately $1\%$, by using the world averages for $\alpha_s$ and the other CKM elements \cite{dEnterria:2016rbf,dEnterria:2020cpv}.

In this work we project the sensitivity of FCC-ee to $|V_{cs}|$ and $|V_{cb}|$, which are the CKM elements for which a substantial improvement, compared to the current precision, can be expected thanks to the $b$, $c$, and $s$ taggers.
Currently, the measurements of these two matrix elements arise from the study of leptonic and semileptonic decays of $D$ and $B$ mesons. 
The most precise measurement of $|V_{cs}|$ stems from $D \to K \ell \nu$ decays, where the dominating uncertainty comes from the lattice QCD calculation of the hadronic form factors. 
The $|V_{cb}|$ matrix element can be extracted either from inclusive $B\to X_c \ell \nu$ or exclusive $B\to D^{(*)} \ell \nu$ semileptonic decays. These two determinations currently have similar precision but are in tension with each other. The uncertainty on $|V_{cb}|$ is already systematically limited and cannot be reduced in the future unless the Belle-II detector performance is better understood \cite{Belle-II:2018jsg}. Regarding $|V_{cs}|$, a precision of approximately $0.2\%$ is expected to be reached at the STCF \cite{Liu:2021qio}.
The current PDG~\cite{ParticleDataGroup:2022pth} averages for $|V_{cs}|$ and $|V_{cb}|$ are reported in the first column of Table~\ref{tab:CurrentVSResults}.

\begin{table}[t]
\renewcommand{\arraystretch}{1}
\centering
\begin{tabular}{cccccc}
\hline\hline
 \multirow{2}{*}{$|V_{ij}|$} & Current & ~ & FCC-ee & FCC-ee  & FCC-ee  \\ 
    & (PDG) & ~ & ($\delta_\epsilon = 1\%$)  & ($\delta_\epsilon = 0.1\%$) & (Stat. only) \\ \hline
$|V_{cs}|$ & $ 0.975 \pm 0.006$ &  $(0.6 \%)$   &  $0.36 \%$ &  $0.05 \%$ &  $0.008 \%$ \\ 
$|V_{cb}|$ & $(40.8 \pm 1.4)\times 10^{-3}$ & $(3.4\%)$   &  $0.52 \%$  & $0.16\%$ & $0.14\%$ \\
\hline\hline
\end{tabular}
\caption{The first column shows the current values of $|V_{cs}|$ and $|V_{cb}|$ from PDG~\cite{ParticleDataGroup:2022pth}. The second, third and fourth columns show the relative precision projected for FCC-ee by assuming 1\%, 0.1\% and zero relative systematic uncertainty on the tagger parameters, respectively.} 
\label{tab:CurrentVSResults}
\end{table}

Early studies \cite{Charles:2020dfl} estimated, from the projected statistics, a relative precision of approximately $0.4\%$ on $|V_{cb}|$, using ILD jet tagging performances as reference. This was later revised to $0.15\%$ using FCC-ee performances \cite{Monteil:VcbWW2024}. Our work follows the same direction, improving the previous ones by discussing possible sources of background and considering the effect of systematic uncertainties on tagger performances. Our results for the expected sensitivity from $WW$ decays, for two different values of relative systematic uncertainty $\delta_\epsilon = 1\%$ and $0.1\%$ and for statistical uncertainty only, are summarized in Table~\ref{tab:CurrentVSResults} and compared with current values. The statistical-only precision for $|V_{cb}|$ is slightly better than the reported value in Ref.~\cite{Monteil:VcbWW2024} in spite of imperfect tagging due to three times more $WW$ pairs being considered.
Such a measurement of CKM matrix elements would be completely independent from those obtained from meson decays. In particular, these will not require any lattice QCD inputs and could provide a final answer to the longstanding tension between inclusive and exclusive determinations of $|V_{cb}|$.

The manuscript is organized as follows. In Section~\ref{sec:Signal} we describe the desired signal and the respective branching fractions, while in Section~\ref{sec:Background} we study possible backgrounds. Next, in Section~\ref{sec:ProbModel} we describe the probabilistic model to extract the FCC-ee sensitivity, including all systematic uncertainties. Finally, in Section~\ref{sec:Results} we present the results and discuss the effect of different tagger parameters on the projected reach.

\section{Signal}
\label{sec:Signal}
The measurement of CKM elements at future lepton colliders is based on counting the number of jets with specific flavors produced by a $W$ decay. At center of mass \mbox{$\sqrt{s} \approx 162$~GeV}, the FCC-ee total planned luminosity will amount to $N_{WW} \approx 3\times10^8$ of $W^+ W^-$ pairs. Each $W$ is emitted approximately at rest and decays promptly to two quark jets or to a charged lepton and a neutrino, with branching ratios ${\rm Br}(W\to{\rm had}) \equiv \mathcal{B}_{\rm had} \simeq 67.4\%$ and ${\rm Br}(W\to\ell\nu) \simeq 32.6\%$ respectively~\cite{ParticleDataGroup:2022pth}.

The decay width of the process $W^+\to u_i \bar d_j$ is
\beq
\Gamma(W^+\to u_i \bar d_j) = 3 |V_{ij}|^2 \Gamma_0 \equiv \Gamma^+_{ij}\,,
\eeq
where at leading order (LO) $\Gamma_0 \approx g_2^2 m_W / (48\pi)$, $g_2$ is the ${\rm SU}(2)_L$ coupling constant, $m_W = 80.4$~GeV is the $W$ boson mass, and the factor of 3 comes from the number of quark colors.
The indices run over the kinematically allowed quark flavors: $i = 1,2 $ for up-quarks and $j = 1,2,3$ for down quarks.
The same expression is valid for $W^-\to\bar u_i d_j$.

The total hadronic width follows by summing over all indices $ij$, that is
\beq
\Gamma^\pm_{\rm had} = \sum_{ij} \Gamma^\pm_{ij} = 3 \Gamma_0 \sum_{ij} |V_{ij}|^2 = {\cal B}_{\rm had} \Gamma_{\rm tot}\,,
\eeq
where $\Gamma_{\rm tot}$ is the total $W$ boson decay width. Using this, we can rewrite the branching ratio for the flavour set $ij$ as
\beq\label{eq:Bij}
{\cal B}_{ij} = \frac{\Gamma_{ij}^\pm}{\Gamma_{\rm tot}} = \frac{|V_{ij}|^2}{\sum_{lm}|V_{lm}|^2} \mathcal{B}_{\rm had} \, ,
\eeq
where we take the total hadronic branching ratio from the PDG~\cite{ParticleDataGroup:2022pth} combination of LEP and CMS results; note however that, at the time FCC-ee or other future circular lepton colliders will be running, the best measurement will be obtained from that same collider as well.

We consider as signal events the processes
\beq
e^+e^-\to W^+W^-\to \lp u_i \bar d_j \rp \lp d_k \bar u_z \rp\,,
\eeq
that is where both $W$ decay hadronically. 
The expected fraction of events per channel is then
\beq\label{eq:Fractionij}
F_{ij} = {\cal A}_W \times {\cal B}_{ij}\,,
\eeq
where ${\cal A}_W$ indicates the total detector acceptance for an hadronic $W$ decay. Finally, the expected number of events for a flavor combination $ijkz$ is
\beq\label{eq:ExpectedNumberij}
N_{ijkz} = N_{WW} \times F_{ij}\times F_{kz}\,.
\eeq

The CKM element $|V_{ij}|$ can be measured by counting the number of decays into the final state $\lp u_i d_j \rp$, where we are considering both charge combinations. In order to cancel systematic uncertainties associated with the total rate, the probabilistic model introduced in Section~\ref{sec:ProbModel} is defined in terms of single-event probabilities per final state and thus constrains $|V_{ij}|$ through the ratio
\beq\label{eq:BijBhad}
\frac{{\cal B}_{ij}}{\mathcal{B}_{\rm had}} =\frac{\Gamma^\pm_{ij}}{\sum_{lm} \Gamma^\pm_{lm}}  = \frac{|V_{ij}|^2}{\sum_{lm}|V_{lm}|^2}.
\eeq
where the denominator is computed using the PDG~\cite{ParticleDataGroup:2022pth} reported best-fit values for all the CKM matrix elements that we are not leaving free in the fit. In each final state $\lp u_i d_j \rp$, the couple of jets can be produced by either one $W$ boson, while the other produces a different pair of jets, or by both $W$ bosons at the same time. The efficiency in each final state depends on the performance of the jet flavor taggers, which are discussed in Section \ref{sec:ProbModel}.

QCD corrections can affect the final $4q$ state in several ways. Higher-order corrections to the  $W \to q \bar{q}^\prime$ decay cancel, up to negligible kinematical effects, when we write the branching ratio ${\cal B}_{ij}$ in terms of the measured total hadronic branching ratio as in Eq.~\eqref{eq:Bij} \cite{dEnterria:2016rbf}. On the other hand, QCD interactions between the two pairs of quarks can induce colour reconnection effects, affecting the final hadron distribution \cite{Gustafson:1988fs,Sjostrand:1993hi,ALEPH:2013dgf}. This phenomenon takes place during the fragmentation process and can be modelled by modern showering algorithms \cite{Christiansen:2015yca}, which is a requirement for a precise measurement of the $W$ boson mass.
Since the goal of this work is to provide an estimate of the sensitvity reach in the CKM matrix element extractions, in the following we work at the parton level doing a simple counting analysis and we assume that any uncertainties due to the colour reconnection modelling can be embedded in the systematic uncertainties of the jet flavour tagging efficiencies.

Finally, we note that the semileptonic channel, $e^+e^-\to WW\to \lp \ell\nu \rp \lp u_i d_j \rp$, could also be exploited for the CKM measurement. The main advantage would be the absence of Drell-Yan backgrounds, which we discuss in the next Section, and of colour reconnection effects. On the downside, the smaller branching ratio in leptons leads to a factor of $\sim1/2$ in the total number of hadronic events, while the invisible neutrino in the final state might pose other experimental challenges. Nevertheless, the semileptonic channel can be employed to expand the present study, or to control the tagger efficiencies in the $WW$ runs. We leave these considerations for future works and proceed here to consider the purely hadronic decays.

\section{Background}
\label{sec:Background}

\begin{table}[t]
\renewcommand{\arraystretch}{1}
\centering
\begin{tabular}{c | c c c c c c c c c c}
\hline\hline
    $W_1$     & $(\bar{d} d)$ & $(\bar{s} s)$ & $(\bar{b} b)$ & $(\bar{u} u)$ & $(\bar{c} c)$ & $(\bar{d} g)$ & $(\bar{s} g)$ & $(\bar{b} g)$ & $(\bar{u} g)$ & $(\bar{c} g)$ \\
    $W_2$     & $(g g)$ & $(g g)$ & $(g g)$ & $(g g)$ & $(g g)$ & $(d g)$ & $(s g)$ & $(b g)$ & $(u g)$ & $(c g)$ \\ \hline
$\sigma$ [fb] & 14 & 14 & 6.9 & 20 & 22 & 15 & 15 & 8.7 & 19 & 22 \\
    $N_{\rm ev}$ [$10^5$] & 1.7 & 1.7 & 0.83 & 2.4 & 2.7 & 1.8 & 1.8 & 1.1 & 2.3 & 1.6 \\
    \hline\hline
\end{tabular}
\caption{Detailed composition of the flavour content of background Drell-Yan events of type $2q2g$, after the $m_W$ invariant mass cut and paired accordingly, with the corresponding cross section.} 
\label{tab:bkg_flavour}
\end{table}

The main background to the $e^+ e^- \to W^+ W^- \to 4 j$ process is the Drell-Yan induced one, $e^+ e^- \to q \bar{q} + 2 j$, where the two extra jets can be either both gluons or a quark anti-quark pair. 
To estimate its importance we simulate $10^6$ Drell-Yan (DY) induced events at LO with {\tt MadGraph5\_aMC}~\cite{Alwall:2011uj}, restricting for simplicity to parton level. We fix the center of mass energy to $\sqrt{s} = 162$ GeV and impose the following cuts: $p_T^j > 5$ GeV, $\eta_j < 2$, $\Delta R_{jj} > 0.1$, yielding the total cross section $\sigma \approx 22.7$pb. Through the same process, we find that other backgrounds, as the  $e^+ e^- \to Z^*Z^*/\gamma^*\gamma^* \to 4 j$ chain, are negligible.

We apply a simple selection cut to the simulated events, by requiring events in which there exist two pairs of jets with invariant masses $(m_{12}, m_{34})$ within 5 GeV of $m_W$,
consistent with the signal being four jets coming from two $W$ decays. We find that $\sim6\times10^3$ background events pass this cut, with the resulting cross section $\sigma_{4j - m_W} \approx 159$ fb. The latter can be decomposed in two categories: final state with two quarks and two gluons, with $\sigma_{2q2g} \approx 158$ fb, and final state with four quarks, with $\sigma_{4q} \approx 0.95$ fb.
The vast majority of the brackground events thus falls in the first category, while only $\sim1\%$ goes in the second. The detailed flavour decomposition of these background events is reported in Table~\ref{tab:bkg_flavour}.
To derive a corresponding number of events, the reported cross section should be multiplied by the integrated luminosity at the $WW$ threshold scan $\mathcal{L}_{WW} \approx 12 {\rm ab}^{-1}$ \cite{FCC:2018evy}. Given the dominance of gluons in the background, we expect an increased rejection rate due to the low mistag probabilities of b-, c-, and s-flavor taggers versus gluon jets, see Table~\ref{tab:JetTaggers}.

\begin{table}[t]
\renewcommand{\arraystretch}{1}
\centering
\begin{tabular}{ccccccc}
\hline\hline
 ~~~ & ~$b$~ & ~$s$~ & ~$c$~ & ~$u$~ & ~$d$~ & ~$g$~ \\ \hline 
$\epsilon_\beta^b$ & 0.8 & 0.0001 & 0.003 & 0.0005 & 0.0005 & 0.007 \\ 
$\epsilon_\beta^c$ & 0.02  & 0.008 & 0.8 & 0.01 & 0.01 & 0.01\\ 
$\epsilon_\beta^s$ & 0.01 & 0.9 & 0.1 & 0.3 & 0.3 & 0.2 \\\hline\hline
\end{tabular}
\caption{Jet-flavor taggers Working Points, indicated by the probabilities $\epsilon_\beta^q$ to tag a $\beta$-jet as a $q$-jet, with $q=b,c,s$ and $\beta = b,s,c,u,d,g$ (see text for details). Values are taken from Refs.~\cite{Bedeschi:2022rnj, Gouskos:taggers, Selvaggi:taggers}. 
} 
\label{tab:JetTaggers}
\end{table}

\section{Probabilistic model and systematics}
\label{sec:ProbModel}
We present the probabilistic model for measuring $|V_{cb}|$ with two orthogonal taggers in order to introduce the notation in a simple way, then extend the same to the general $|V_{ij}|$ with two or three orthogonal taggers.

The Parameter of Interest (POI) in our projection is $\mu = |V_{cb}|$. As the true value we take the PDG~\cite{ParticleDataGroup:2022pth} central value: $\mu_{\rm true} = |V_{cb}|_{\rm PDG}$.
Following the method detailed in Refs.~\cite{ATLAS:2022ers, Faroughy:2022dyq, CMS:2020vac}, we apply $b$- and $c$-taggers\footnote{We are assuming here that the two taggers are orthogonal to avoid double-countings, e.g. see Refs.~\cite{Faroughy:2022dyq,Kamenik:2023hvi}} to the final state jets and separate each couple in tag bins $(n_b,n_c) = \{ (0,0),(1,0),(0,1),(2,0),(0,2),(1,1)\}$. Here $n_b$ and $n_c$ indicate the number of jets tagged as $b$ and $c$ respectively; that is, we are counting not only events where the jets are tagged correctly, falling into bin $(1,1)$, but also those where one or both jets are mistagged.
With fully hadronic $W_1 W_2$ events, the expected number of events in the two pairs of bins, $B_{bc;1}\equiv(n_{b;1},n_{c;1})$ and $B_{bc;2}\equiv(n_{b;2},n_{c;2})$, is then
\beq
N_{B_{bc;1},B_{bc;2}}(\mu,\nu) = N_{WW} \prod_{i=1,2}\sum_f p(n_{b;i},n_{c;i}|f,\nu) F_f(\mu,\nu) \,,
    \label{eq:Nevents_bins}
\eeq
where the product goes over the two bosons, while the sum spans over all the possible final states for a $W$, $f=\{ ud, us, ub, cd, cs, cb \}$.
Since we assume that the taggers cannot distinguish quarks from antiquarks, $N_{B_{bc;1},B_{bc;2}}$ is a $6\times 6$ symmetric matrix. The off-diagonal elements can be merged, providing 21 different event categories, however for simplicity we keep them separate in the discussion since this does not affect the final result.
The expected fraction of events for $f$, $F_{f}$ (see eq.~\eqref{eq:Fractionij}), depends on the POI and on a set of nuisance parameters $\nu$, described below.
The function $p(n_{b;i},n_{c;i}|f,\nu)$ is the probability distribution of an event with final state $f$ to fall into the $(n_{b;i},n_{c;i})$ bin. We provide more details on this function in Appendix~\ref{app:probfunc}. Note again that this function depends on both the final state and the nuisance parameters.

\begin{table}[!t]
\renewcommand{\arraystretch}{1}
\centering
\begin{tabular}{ccc}
\hline\hline
 ~~~~Parameter~~~~ & ~~~~Value~~~~ \\ \hline 
$N_{WW}$ & $3\times10^8$ \\ 
${\rm Br}(W\to{\rm had})$ & $0.6741$  \\ 
${\cal A}_W$ & 0.9  \\
\hline\hline
\end{tabular}
\caption{Fixed parameters from physics and detector inputs considered in the fit.
} 
\label{tab:NuisanceParameters}
\end{table}

The nuisance parameters $\nu$ include physics and detector inputs, like branching ratios and acceptances, as well as tagger performance parameters. We list the former in Table~\ref{tab:NuisanceParameters}, showing their numerical values; their relative uncertainties are small enough~\cite{FCC:2018byv} that their effect on the fit is negligible, thus we consider them as fixed inputs. The latter are shown in Table~\ref{tab:JetTaggers}: for the $q$-tagger, we denote as $\epsilon_\beta^q$ the probability to tag a $\beta$-jet as a $q$-jet, where $\beta=\{ b,s,c,u,d,g\}$.
Currently, the level of systematic uncertainty on these parameters, $\delta_\epsilon\equiv\delta\epsilon/\epsilon$, is of the order of few percent~\cite{ATLAS:2019bwq, CMS:2017wtu}, while future colliders will be able to reduce it to ${\cal O}(1)\%$. Recent studies on $b$-taggers at FCC-ee~\cite{Selvaggi:Rb} show that a systematic of ${\cal O}(0.1)\%$ is achievable, in particular thanks to the dedicated calibration during the running at the $Z$-pole. In order to show the effect of the systematic uncertainties on $\epsilon_\beta^q$ in our results, we fix three benchmarks for our numerical analysis: $\delta_\epsilon=0,~0.1,~1$ \%. In each of these cases, we assume for simplicity that $\delta_\epsilon$ is the same for all tagger parameters.

Given the discussion in Section~\ref{sec:Background}, we are also assuming that the DY background can be effectively reduced to a negligible fraction, thus not entering in the definition of $N_{B_{bc;1},B_{bc;2}}(\mu,\nu)$.\footnote{We checked that including DY events in the case $\delta_\epsilon = 0$ has a negligible impact in the sensitivity to $|V_{cs}|$ and $|V_{cb}|$. An even smaller effect is expected when systematic uncertainties are included.} 
As a further simplification in our analysis, we exploit the large statistics expected in each tag bin $B_{bc;i}$ to bypass the need of detailed Monte Carlo signal simulations, and instead work in the Asimov approximation~\cite{Cowan:2010js}. In the Asimov approximation, one sets the observed values of $N_{B_{bc;i}}$ to $N^A_{B_{bc;i}} = N_{B_{bc;i}}(\mu=\mu_{\rm true},\nu=\nu_0)$, where $\mu_{\rm true}$ is the input value of $\mu$ (the PDG central value in this work) and $\nu_0$ is the nominal value of each nuisance parameter.

With the above ingredients, we can build the likelihood of the observed number of events, $N^A_{B_{bc;1},B_{bc;2}}$, given the expected number of events $N$, see Eq.~\eqref{eq:ExpectedNumberij}, as
\beq
{\cal L}(\mu,\nu) = {\cal P}( N^A_{B_{bc;1},B_{bc;2}}|N_{B_{bc;1},B_{bc;2}}(\mu,\nu) )p(\nu)\,,
    \label{eq:likelihood}
\eeq
where $\mathcal{P}$ is the Poisson likelihood
\beq
{\cal P}(k|x) = \frac{x^k e^{-x}}{k!}\,,
\eeq
and $p(\nu)$ is the appropriate distribution for the nuisance parameters, which we assume to be a normal distribution, centered on the nominal value and with the given uncertainty as the standard deviation. We follow Ref.~\cite{Cowan:2010js} and define the profile likelihood ratio
\beq\label{eq:plr}
    \lambda(\mu)= \frac{{\cal L}(\mu,\hat{\hat{\nu}}(\mu))}{{\cal L}(\hat{\mu},\hat{\nu})}\,,
\eeq
with the associated test statistic
\beq\label{eq:test_stat}
   t_{\mu}= -2\ln\lambda(\mu)\,.
\eeq 
Here, $\hat{\hat{\nu}}(\mu)$ are the maximum likelihood estimates (MLE) of the nuisance parameters, obtained by  maximizing $\mathcal{L}({\mu},{\nu})$, varying $\nu$, but keeping $\mu$ fixed. The MLE $\hat{\mu}$ and $\hat{\nu}$ are instead obtained by finding the global maximum of $\mathcal{L}(\mu,\nu)$, varying both $\nu$ and $\mu$. In the Asimov approximation, the MLE are by definition $\hat{\mu}=\mu_{\rm true}$ and $\hat{\nu}=\nu_0$. 

Finally, we can extract the confidence intervals on $\mu$.
In particular, the $68\%$ confidence interval $[\mu_{\rm low},\mu_{\rm up}]$ is obtained by solving for $t_{\mu} = 1$; given the large number of events, we can practically approximate the latter by a $\chi^2$ distribution.
We report our results as $\delta|V_{cb}|/|V_{cb}|$, that is the relative uncertainty on the determination of the CKM matrix element.

The model described in this Section can be trivially generalized for all CKM elements, or extended to fit simultaneously more than one variable. In the case of interest here, we can perform a combined fit of $|V_{cs}|$ and $|V_{cb}|$ by defining two POI, $\mu_{cs}$ and $\mu_{cb}$, and extend the probability distribution function to be defined over 3-dimensional bins. Namely, we separate events in the tag bins $(n_b,n_c,n_s)$ by means of the function $p(n_b,n_c,n_s | f,\nu)F_{f}(\mu_{cs},\mu_{cb},\nu)$ and compute the corresponding $68\%$ CL contours by solving for $t_{\mu_{cs},\mu_{cb}}=2.28$. We provide details on $p(n_b,n_c,n_s | f,\nu)$ in Appendix~\ref{app:probfunc}.

\section{Results}
\label{sec:Results}

\begin{figure}[t]
\begin{center}
\includegraphics[width=0.75\linewidth]{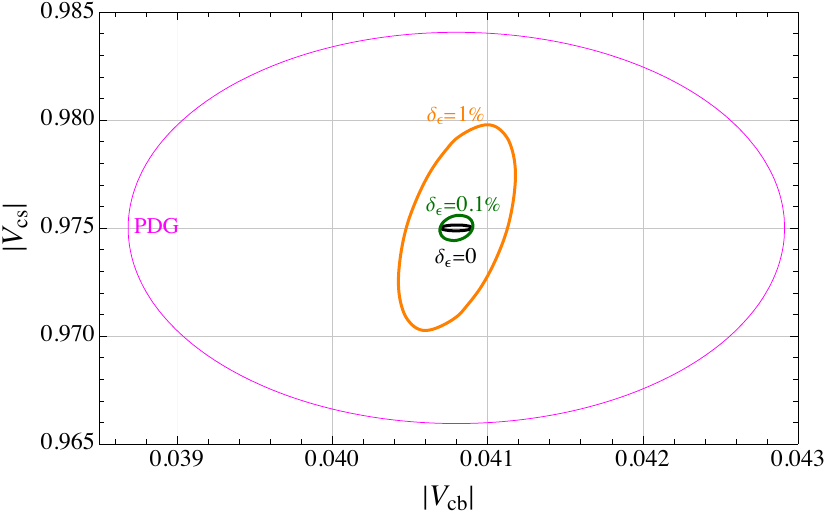}
\end{center}
\vspace{-0.3cm}
\caption{68\% CL contours in the $(|V_{cb}|,|V_{cs}|)$ plane. The magenta contour corresponds to the present precision~\cite{ParticleDataGroup:2022pth}, while orange, green, and black contours show the expected sensitivity from $W$ decays assuming 1\%, 0.1\% and 0 systematic uncertainties on the tagging efficiencies, respectively.}
\label{fig:Vcb_Vcs_sensitivity}
\end{figure}

Here we apply our method to study the cases of $|V_{cb}|$ and $|V_{cs}|$. The main results are summarized in the last three columns of Table~\ref{tab:CurrentVSResults} and are obtained by considering one matrix element at a time and only the two more relevant taggers, whose efficiencies are set to the values shown in Table~\ref{tab:JetTaggers}. The second and third columns correspond to two benchmarks for systematic uncertainties, $\delta_\epsilon = 1\%,~0.1\%$ and the last column corresponds to the statistical uncertainty only scenario. In all cases, FCC-ee is able to improve on the current levels of precision. Here and in the following, we only consider systematic uncertainties for the tagger efficiencies. We have found that the other systematic uncertainties are subleading and do not qualitatively alter our results.

In Fig.~\ref{fig:Vcb_Vcs_sensitivity} we show the 68\%~CL contours in the plane of $|V_{cb}|$ and $|V_{cs}|$, by performing a combined analysis with the three taggers for different values of $\delta_\epsilon$ (black, green and orange lines), compared to the present constraints obtained from PDG by neglecting possible correlations (magenta line). The combined analysis needs to assume that the three taggers are orthogonal, which may be unjustified for current state-of-the-art taggers. However, ongoing efforts using multi-class algorithms will produce such a set of orthogonal taggers with high performance, see e.g. Ref.~\cite{Bedeschi:2022rnj, Gouskos:taggers, Selvaggi:taggers,Liang:2023yyi}.

\begin{figure}[t]
\begin{center}
\includegraphics[width=0.75\linewidth]{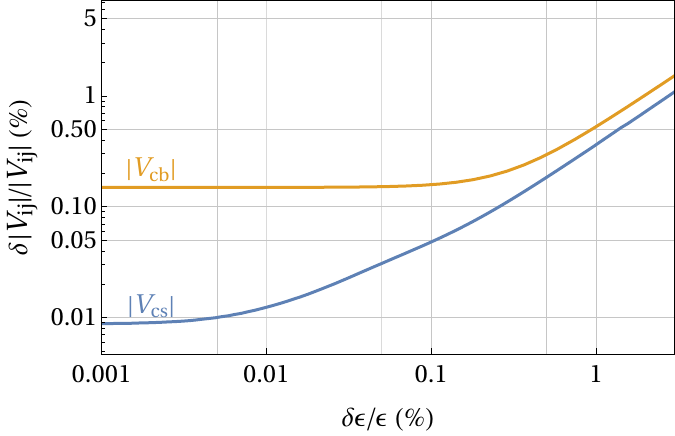}
  \vspace{-0.3cm}
\end{center}
\caption{Sensitivity reach on $|V_{cs}|$ (blue) and $|V_{cb}|$ (orange) as function of the systematic uncertainty on the $\epsilon$ parameters.
}\label{fig:syst_scan}
\end{figure}

Since Fig.~\ref{fig:Vcb_Vcs_sensitivity} indicates that there is only a mild correlation between the two CKM elements (with a correlation coefficient of $\approx 0.5$ for $\delta_\epsilon =1\%$), we can analyze $|V_{cb}|$ and $|V_{cs}|$ separately using two taggers at a time and directly compare the one-dimensional confidence intervals with those from PDG.
In Fig.~\ref{fig:syst_scan} we show the projected relative sensitivity on the two CKM matrix elements as a function of the systematic uncertainty.
The role of systematics is immediately clear in both cases, albeit with different behaviours. For $|V_{cb}|$ (orange line), the precision degrades by a factor of 3 if we switch from the best scenario to the 1\% systematic level case. Nevertheless, even with the latter, a precision on $|V_{cb}|$ of about $0.5\%$ is achievable, which is already an order of magnitude smaller than current results, see Table~\ref{tab:CurrentVSResults}. However, at systematics of ${\cal O}(0.1\%)$, the result saturates to a constant and stops improving with even lower uncertainties. This is due to the relatively small number of events in the $cb$ decay channel. That is, statistical uncertainties on $|V_{cb}|^2$ start to dominate the error budget, and further improvements on the tagger side do not help.
Differently, for $|V_{cs}|$ (blue line) the result does not flatten at the 0.1\% level but keeps decreasing. Unlike the previous case, the very large statistics in the $cs$ channel renders the measurement systematic-dominated for any realistic value of systematic uncertainties.

\begin{figure}[t]
\begin{center}
\includegraphics[width=0.9\linewidth]{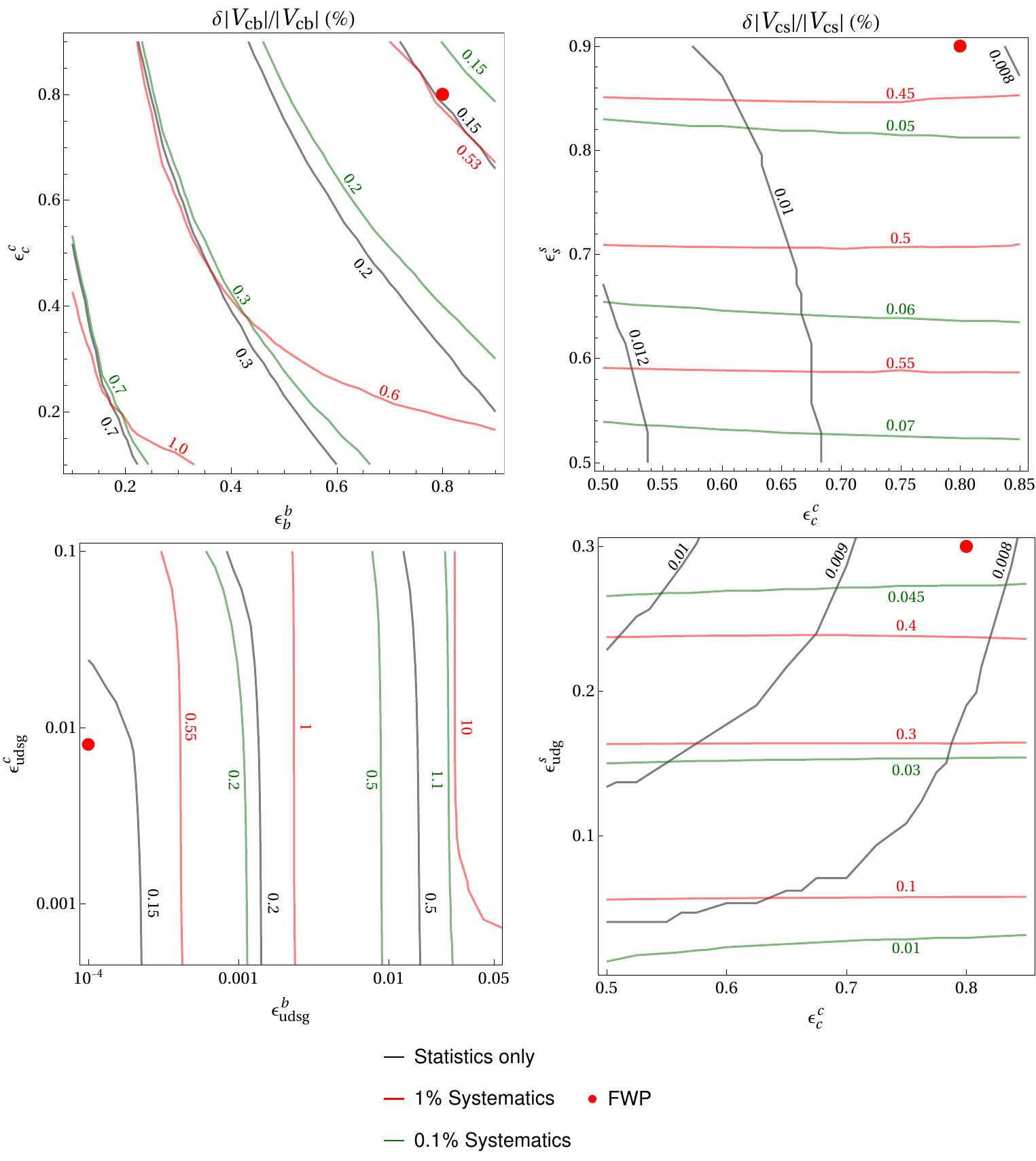}
  \vspace{-0.3cm}
\end{center}
\caption{Precision contours on $|V_{cb}|$ (left) and $|V_{cs}|$ (right) obtained by scanning over tagger parameters. The black, red and green lines indicate the results with statistical uncertainty only, 1\% and 0.1\% systematics uncertainties respectively, while the red dot corresponds to the values of the working point, 
see Table~\ref{tab:JetTaggers}. {\bf Left column}: precision contours on $|V_{cb}|$ in the  $\{ \epsilon_b^b, \epsilon_c^c \}$ (top) and  $\{ \epsilon_b^{udsg}, \epsilon_c^{udsg} \}$ (bottom) planes.  {\bf Right column}: precision contours on $|V_{cs}|$ in the $\{ \epsilon_c^c, \epsilon_s^s \}$ (top) and  $\{ \epsilon_c^c, \epsilon_s^{udg} \}$ (bottom) planes.
}\label{fig:Vcb_Vcs_tagger_contours}
\end{figure}

We now move to study the dependence of our results on the specific working point for the tagging efficiencies, for the three benchmarks of the systematic uncertainties.
In Fig.~\ref{fig:Vcb_Vcs_tagger_contours} (left column) we show isocontours of $\delta|V_{cb}|/|V_{cb}|$ by varying different combinations of tagger parameters. In both plots, we show with black, red and green lines the results with the three different uncertainty assumptions, statistical uncertainty only, 1\% and 0.1\% systematics respectively, while the red dot indicates the tagger working point values. 
In the top plot, we scan the $\{ \epsilon_b^b, \epsilon_c^c \}$ plane. The results present only a mild dependence on these parameters: to drop the precision by a factor of 2 for small systematics, the efficiencies of both taggers need to be halved. The same corresponds to a ${\cal O}(10)\%$ drop for the 1\% benchmark. 

Given the result discussed above and the relatively small statistics of $cb$ final states, the main effect on the precision will come from mistagging a light jet as a $c$- or a $b$-jet.
Thus, in the bottom plot we turn our attention to the $\{ \epsilon^b_{udsg}, \epsilon^c_{udsg} \}$ plane. It is evident in this scan how a powerful rejection of light jets from the $b$-tagger, together with low systematics, are necessary to obtain the desired sensitivity. While projected performances and uncertainties are promising, it is worth noting that $\epsilon_{udsg}^b = 0.005$ and $\delta\epsilon/\epsilon = 1\%$ will bring the FCC-ee sensitivity on $|V_{cb}|$ down to the same order as the current levels. On the other hand, the requirements on the $c$-tagger performance can remain relatively loose, given the very large number of $c$ quarks produced.

Regarding $|V_{cs}|$, the large effect from systematic uncertainties on $\epsilon$, observed in Fig.~\ref{fig:syst_scan}, is evident in Fig.~\ref{fig:Vcb_Vcs_tagger_contours} (right column), where we show isocontours of $\delta|V_{cs}|/|V_{cs}|$ by again varying different combinations of tagger parameters. 
In the top plot, we show the percentual precision as a function of the two relevant true positive rates, $\{ \epsilon_c^c, \epsilon_s^s \}$. For statistical uncertainty only, the behaviour is fairly straightforward. Increasing either efficiency increases the precision, and the isocontours are obtained by decreasing $\epsilon_s^s$ as $\epsilon_c^c$ increases. The introduction of systematic uncertainties for the tagger efficiencies not only degrades the overall performance but also changes the qualitative features of the scans:
the isocontours are almost independent on $\epsilon_c^c$.
This indicates a relatively large lack of sensitivity to changes in $\epsilon_c^c$, with $\epsilon_s^s$ driving the precision. A similar behavior is observed in the bottom plot, which shows the scan on the $\{ \epsilon_c^c, \epsilon_{udg}^s \}$ space.

To understand the flattening of performance curves in the $\epsilon_c^c$ direction, we need to look at the nuisance parameters of the fit. By adding systematic uncertainties, we need to take into account the fact that the nuisance parameters can become constrained by the measurements, at the expenses of precision on $|V_{cs}|$, and will have a particular correlation structure, which in turn will affect how each decay channel populates different bins. On the $c$-tagger side, the mistag parameter $\epsilon_{udg}^{c}$ is almost negligible, as well as backgrounds with possible false $c$-tags originating from $b$-jets, given the hierarchical structure of the CKM matrix
($|V_{ub}|,|V_{cb}|\ll |V_{us}|,|V_{cd}|\ll |V_{ud}|,|V_{cs}|$).
On the $s$-tagger side, the nuisance parameters will impact the fit mainly by modifying the different sources of $s$-tags. These nuisance parameters become correlated by the statistical fit and are constrained mainly by the low signal bins with well-known backgrounds, which we are keeping fixed here. A change in one of the $\epsilon_\beta^s$ efficiencies will modify the signal-to-background ratio for all bins and affect the constraints on the nuisance parameters, impacting the $|V_{cs}|$ precision, while a change in $\epsilon_c^c$ can be almost compensated by a correlated change in all $\epsilon_\beta^s$. Additionally, because $\epsilon_{udg}^{s}$ is particularly large compared to other mistags, its absolute uncertainty becomes larger. 

\section{Conclusions}

In this work we showed the potential of FCC-ee, or CEPC, to measure $|V_{cs}|$ and $|V_{cb}|$ from hadronic $W$ decays.\footnote{Our results could also be applied to ILC by choosing a different working point for the tagging efficiencies in Fig.~\ref{fig:Vcb_Vcs_tagger_contours}.}  Our findings are summarized in Table~\ref{tab:CurrentVSResults}. In both cases, the relative uncertainty will largely surpass present levels.
Assuming per-mille level systematics on the tagger parameters, one can expect to increase the precision on CKM elements by one order of magnitude.
More conservative assumptions for the tagger performances will lead to smaller improvement, especially in the systematic dominated case of $|V_{cs}|$. Together with the systematic level, final sensitivities strongly depend on the rejection capability of the taggers, that is on the mistag probabilities. Projections on tagger working points at FCC-ee indicate that such low mistag rates can be achieved; particular attention will be needed for $s$-tagger, where large $s$-to-light jet mistagging can sizeably degrade the results. 

The knowledge of CKM matrix elements is often the limiting factor in the sensitivity to physics beyond the SM from rare decays and meson mixing, which would therefore benefit from such high precision \cite{Charles:2020dfl}.
This measurement will also mark the first \emph{lattice-free} determination of $|V_{cb}|$, which will be able to resolve the tension between inclusive and exclusive results from meson decays.
Furthermore, such a precise measurement of CKM elements would allow to use them as inputs in the prediction of hadron decays. Assuming absence of new physics in leptonic or semi-leptonic charged-current decays this could facilitate the direct measurement of meson decay constants and hadronic form factors, providing a benchmark for comparisons with lattice QCD results.

Additional advancements in tagger performances can open the way to measure other relevant quantities; e.g., employing a $d$-tagger~\cite{Selvaggi:taggers} as an anti-light jet tagger, it would be possible to measure $|V_{cd}|$ with the same procedure described in this paper, achieving a precision comparable or superior to present results. We leave these considerations for future work.

\section*{Acknowledgments}
The authors would like to thank Jernej Kamenik, LingFeng Li, Stephane Monteil, Marie-Hélène Schune, Michele Selvaggi and Jure Zupan for useful discussions. MS acknowledges support in part by the DOE grant de-sc0011784 and NSF OAC-2103889. DM acknowledges
support by the MUR grant PRIN 20224JR28W.

\begin{appendix}

%
\section{Details on the probability function}
\label{app:probfunc}
%
Here we provide further details on the probability function introduced in Section~\ref{sec:ProbModel}.

The probability distribution function $p(n_{b},n_{c}|f,\nu)$ is the probability for a particular decay, with true parton flavor configuration $f = u_i, d_j$, to end up in the bin tagged as $(n_b, n_c)$, with $n_b + n_c \leq 2$.
These probabilities are built in terms of the jet tagging efficiencies: $\epsilon^q_{\beta}$ is the probability that a jet of true flavour $\beta$ is tagged as a $q$-jet, where $q$ can be $b$ or $c$. The probability of a $\beta$-jet to not be tagged as neither of the two is therefore $(1 - \epsilon^c_\beta - \epsilon^b_\beta)$.

From these, we define the probabilities to have $W$ events in each $(n_{b}, n_{c})$ bin, summing over the hadronic final states $f$ as in Eq.~\eqref{eq:Nevents_bins}:
\beq
    P_W(n_{b}, n_{c}) \equiv \sum_f p(n_{b},n_{c}|f,\nu) F_f(\mu,\nu).
\eeq
Explicitly, the 6 probabilities are given by
\beq\begin{split}
    P_W(1,1) &= \sum_{u_i = u, c} \sum_{d_j = d,s,b} (\epsilon^c_{u_i} \epsilon^b_{d_j} + \epsilon^b_{u_i} \epsilon^c_{d_j}) {\cal B}_{ij} {\cal A}_W~, \\
    P_W(0,2) &= \sum_{u_i = u, c} \sum_{d_j = d,s,b} \epsilon^c_{u_i} \epsilon^c_{d_j} {\cal B}_{ij} {\cal A}_W~, \\
    P_W(2,0) &= \sum_{u_i = u, c} \sum_{d_j = d,s,b} \epsilon^b_{u_i} \epsilon^b_{d_j} {\cal B}_{ij} {\cal A}_W~, \\
    P_W(0,1) &= \sum_{u_i = u, c} \sum_{d_j = d,s,b} (\epsilon^c_{u_i} (1-\epsilon^c_{d_j} - \epsilon^b_{d_j}) + \epsilon^c_{d_j} (1-\epsilon^c_{u_i} - \epsilon^b_{u_i}) ) {\cal B}_{ij} {\cal A}_W~, \\
    P_W(1,0) &= \sum_{u_i = u, c} \sum_{d_j = d,s,b} (\epsilon^b_{u_i} (1-\epsilon^c_{d_j} - \epsilon^b_{d_j}) + \epsilon^b_{d_j} (1-\epsilon^c_{u_i} - \epsilon^b_{u_i}) ) {\cal B}_{ij} {\cal A}_W~, \\
    P_W(0,0) &= \sum_{u_i = u, c} \sum_{d_j = d,s,b} (1-\epsilon^c_{u_i} - \epsilon^b_{u_i})(1-\epsilon^c_{d_j} - \epsilon^b_{d_j}) {\cal B}_{ij} {\cal A}_W~,
\end{split}\eeq
where ${\cal B}_{ij}$ are the $W$ branching ratios in each channel defined in Eq.~\eqref{eq:Bij} and ${\cal A}_W$ is the acceptance. It can be explicitly checked that the sum of these 6 probabilities gives precisely ${\cal B}_{\rm had}  {\cal A}_W$, as expected.
Finally, the expected number of events $N_{B_{bc;1},B_{bc;2}}$ is obtained as
\beq
    N_{B_{bc;1},B_{bc;2}} = N_{WW} \, P_W(B_{bc;1}) \, P_W(B_{bc;2})~.
\eeq
Also this model can be easily extended to the three jet tagging categories by modifying $P_W(n_{b;i}, n_{c;i}) \to P_W(n_{b;i}, n_{c;i} , n_{s;i})$, implying that there would be 10 categories for each $W$ decay, rather than 6. Also, the probability of a jet not being tagged should be modified to $(1-\epsilon^c_{\beta} - \epsilon^b_{\beta} - \epsilon^s_{\beta})$.

Given the large number of expected events in all bins, consequence of $N_{WW} \approx 10^8$, the central limit can be applied to obtain accurate results.
In the Gaussian approximation the likelihood of Eq.~\eqref{eq:likelihood} can be written as
\beq
    - 2 \ln {\cal L} = \sum_{B_{bc;1}, B_{bc;2}} \frac{\left( N_{B_{bc;1},B_{bc;2}} - N^A_{B_{bc;1},B_{bc;2}}\right)^2}{N_{B_{bc;1},B_{bc;2}}} + \chi^2_{\rm tag}~,
\eeq
where as nuisance parameters we consider only the tagging efficiencies with
\beq
    \chi^2_{\rm tag} = \sum_{q = b,c} \sum_{\beta = b,s,c,u,d,g} \left( \frac{\epsilon^q_\beta - \hat{\epsilon}^q_\beta }{\delta_\epsilon \hat{\epsilon}^q_\beta} \right)^2~.
\eeq
Here $\hat{\epsilon}^q_\beta$ are the nominal values of the efficiencies as in Table~\ref{tab:JetTaggers} and $\delta_\epsilon$ is the relative uncertainty, that we assume to be the same for all taggers.

\end{appendix}

\bibliographystyle{JHEP}
\bibliography{references}

\end{document}